\documentclass[pra,twocolumn,showpacs,nopreprintnumbers,amsmath,amssymb]{revtex4}
\usepackage{graphicx}
\usepackage{dcolumn}
\usepackage{bm}

\begin{document}

\preprint{}

\title{
Initial state maximizing the nonexponentially decaying 
survival probability \\ for unstable multilevel systems
%
}

\author{Manabu Miyamoto}%
 \email{miyamo@hep.phys.waseda.ac.jp }
\affiliation{%
Department of Physics, Waseda University, 3-4-1 Okubo, Shinjuku-ku, 
Tokyo 169-8555,  Japan 
}%

\date{\today}

\begin{abstract} 
The long-time behavior of the survival probability for unstable multilevel 
systems that follows the power-decay law 
is studied based on the $N$-level Friedrichs model, 
and is shown to depend on the initial population in unstable states. 
A special initial state 
maximizing the asymptote of the survival probability at long times 
is found and examined 
by considering the spontaneous emission process for the hydrogen atom 
interacting with the electromagnetic field. 
 
\end{abstract}


\maketitle



One of the crucial characters of 
unstable systems is the famous exponential-decay law. 
Observations of the law were made 
for many quantum systems, 
and its theoretical description also proved to be attributed to 
the poles 
on the second Riemann sheet 
of the complex energy plane \cite{Nakazato(1996)}. 
However, 
the deviation from the exponential decay law 
was also predicted both for short times and for long times 
\cite{Khalfin(1957)}.  
Indeed, despite an apparent difficulty \cite{Greenland(1988)}, 
a nonexponential decay law at short times was successfully observed 
\cite{Wilkinson(1997)}. 
On the other hand, the long-time deviation has still not been detected, 
even though expected 
for all systems coupled with 
the continuum of the lower-bounded energy spectrum. 
The main reasons behind the matter could be ascribed to 
too small survival probability, 
that is, the component of the initial state 
remaining in the state at long times.

The unstable systems are described by the Friedrichs model 
\cite{Friedrichs(1948),Exner(1985)}, 
which enables us to investigate the time evolution involving such processes as 
the spontaneous emission of photons from the atoms 
\cite{Facchi(1998),Antoniou(2001)} 
and the photodetachment of electrons from the negative ions 
\cite{Rzazewski(1982),Haan(1984),Antoniou(2001),Nakazato(2003)}. 
In the former often 
only the first excited level is counted, 
while other higher ones are neglected, 
and in the latter 
the negative ion is assumed to have only one electron bound state. 
These single-level approximations (SLA) 
could be verified 
as long as the lowest level is quite separate from the higher ones. 
However, 
the multilevel treatment of the model gives us another advantage: 
the choice of coherently superposed initial-states 
extending over various levels. 
In fact, it can yield a variety of temporal behavior 
that is never found in the SLA 
\cite{Frishman(2001),Antoniou(2003),Antoniou(2004),Miyamoto(2003)}. 
Such multilevel effects on temporal behavior 
are still not well studied 
except for Refs. \cite{Frishman(2001),Antoniou(2003),Antoniou(2004),
Miyamoto(2003),Davies(1974)}, 
and much less examined with respect to nonexponential decay at long times.

In the present article, we consider 
the long-time behavior of the survival probability $S(t)$ 
by examining the $N$-level Friedrichs model. 
In particular, restricting ourselves to the weak coupling case, 
we clarify how the asymptote of $S(t)$ depends on the initial states. 
By choosing the initial state localized at the lowest level, 
we look at the SLA from a multilevel treatment. 
Then, the result in the $N$-level model turns out to agree with that in 
the SLA in the weak coupling regime. 
Furthermore, among the various initial states, 
we can find 
a special one that maximizes the asymptote of $S(t)$ at long times. 
Initial states that eliminate the first term 
of the asymptotic expansion of $S(t)$ are also obtained. 
For clarity of discussion, 
we assume all form factors to vanish at zero energy. 
However, the existence of such special initial states is proved 
to be quite general and independent of other details of the form factors.

The $N$-level Friedrichs model 
describes the couplings between the discrete spectrum 
and the continuous spectrum. 
The Hamiltonian of the model is defined by 
\begin{equation}
H=H_0 +\lambda V ,
\label{eqn:2.10}
\end{equation}
where $H_0$ denotes the free Hamiltonian 
\begin{equation}
H_0 = \sum_{n=1}^{N} \omega_{ n} | n \rangle \langle n | 
+ \int_0^{\infty}  d \omega \ \omega | \omega \rangle \langle \omega |,
\label{eqn:2.20}
\end{equation}
and $\lambda V$ the interaction Hamiltonian 
\begin{equation}
V = 
\sum_{n=1}^{N}
\int_0^{\infty}  d \omega \
\left[ v_n^* (\omega ) | \omega \rangle \langle n | +
v_n (\omega ) | n \rangle \langle \omega | \right],
\label{eqn:2.50}
\end{equation}
with the coupling constant $\lambda$. 
The eigenvalues $\omega_{n} $ of $H_0$ 
were supposed not to be degenerate, i.e., 
$\omega_n < \omega_{n^{\prime}}$ for $n < n^{\prime}$. 
%
%
Both $| n \rangle $ and $| \omega \rangle $ are 
the bound and scattering eigenstates of $H_0$, respectively, 
and satisfy the orthonormality condition: 
$\langle n | n^{\prime} \rangle = \delta_{n n^{\prime}}$, 
$\langle \omega | \omega^{\prime} \rangle =\delta (\omega - \omega^{\prime})$, 
and $\langle n | \omega \rangle = 0 $, 
where $\delta_{n n^{\prime}}$ is Kronecker's delta 
and $\delta (\omega - \omega^{\prime})$ is Dirac's delta function. 
They also compose the complete orthonormal system with 
the resolution of identity. 
%
%
In Eq. (\ref{eqn:2.50}), 
$v_n (\omega )$ denotes the form factor characterizing 
the transition between $| n \rangle$ and $| \omega \rangle$. 
%
%
%
In the latter discussion, 
we will simplify the model with the assumption that
the form factor $v_n (\omega )$ is an analytic function 
in a complex domain including the cut $(0, \infty )$, 
and behaves like 
\begin{equation}
v_n (\omega) = 
\left\{ 
\begin{array}{ll}
q_n \omega^{p_n} & ( \omega \rightarrow +0 ) \\
s_n \omega^{-r_n} & ( \omega \rightarrow \infty )
\end{array} 
\right. ,
\label{eqn:6.40}
\end{equation}
where $p_n$ and $r_n$ are the positive constants, 
while $q_n$ and $s_n$ are appropriate ones. 
The small-energy condition ensures that the integral 
$\int_{0}^{\infty}  d \omega 
v_n (\omega ) v_{n^{\prime }}^* (\omega )/\omega$ 
is definite. 
The large-energy condition ensures 
that $\int_{0}^{\infty}  d \omega 
v_n (\omega ) v_{n^{\prime }}^* (\omega )/(z -\omega) $ 
is definite for all complex numbers $z \notin [0, \infty )$. 
Both of the conditions are satisfied by several systems involving 
the spontaneous emission process of photons 
\cite{Facchi(1998),Antoniou(2001)} and 
the photodetachment process of electrons 
\cite{Rzazewski(1982),Haan(1984),Antoniou(2001),Nakazato(2003)}. 
Note that this small-energy condition excludes 
the photoionization processes associated with the Coulomb interaction 
\cite{Wigner(1948)}; 
however, the formulation developed below could be applied 
to those cases.

The initial state $| \psi \rangle $ of our interest 
is an arbitrary superposition 
of the unstable states $| n \rangle $, 
\begin{equation}
| \psi \rangle = \sum_{n=1}^{N}  c_n | n \rangle , 
\label{eqn:6.10}
\end{equation}
where $c_n$'s are complex numbers 
satisfying the normalization condition 
$\sum_{n=1}^{N}  |c_n |^2 =1$. 
Then, the survival probability $S(t)$ of the initial state $| \psi \rangle$, 
that is, the probability of finding the initial state 
in the state at a later time $t$, is defined by $S(t)= |A(t) |^2 $. 
The $A(t)$ denotes the survival amplitude of $| \psi \rangle $, i.e., 
$A(t) =\langle \psi | e^{- i tH} | \psi \rangle $. 
In general, the Hamiltonian (\ref{eqn:2.10}) 
has the possibility of possessing 
not only the scattering eigenstates $| \psi_{\omega}^{(\pm )} \rangle $, 
but also the bound eigenstates \cite{bound-states}. 
We shall here restrict ourselves to studying the decaying part of $A(t)$, 
and merely call it the survival amplitude 
with the same symbol as 
\begin{equation}
A(t)=\int_{0}^{\infty } 
 d \omega e^{- i t\omega } 
|\langle \psi_{\omega }^{(\pm)} | \psi \rangle |^2  . 
\label{eqn:6.20}
\end{equation}
In order to estimate the long time behavior of $A(t)$, 
let us evaluate the scattering eigenstates 
$| \psi_{\omega }^{(\pm )} \rangle$ 
by solving the Lippmann-Schwinger equation, i.e., 
$
| \psi_{\omega }^{(\pm )} \rangle
= | \omega \rangle + (\omega \pm  i  0 -H_0)^{-1} \lambda V 
| \psi_{\omega }^{(\pm )} \rangle
$. 
In the case of our Hamiltonian, 
this equation can be solved 
in the form, 
$
| \psi_{\omega }^{(\pm)} \rangle
=
| \omega \rangle + 
\sum_{n=1}^{N} F_n^{(\pm)} (\omega ) 
\left[ 
| n \rangle + 
\int_0^{\infty }  d \omega^{\prime } 
\frac{\lambda v_n^* (\omega^{\prime} )}{\omega - \omega^{\prime} \pm i  0} 
| \omega^{\prime} \rangle 
\right] ,
$
from which the integrand of $A(t)$ reads,
\begin{equation}
\langle \psi_{\omega }^{(\pm)} | \psi \rangle 
=\sum_{n=1}^{N} F_n^{(\pm) *} (\omega ) c_n .
\label{eqn:220}
\end{equation}
The $F_n^{(\pm)} (\omega )$ is determined by an algebraic equation 
\begin{equation}
\sum_{n^{\prime} =1}^{N} G^{-1}_{n n^{\prime }} (\omega \pm i0) 
F_{n^{\prime}}^{(\pm)} ( \omega ) =-\lambda v_n (\omega ) , 
\label{eqn:3.50}
\end{equation}
where
\begin{equation}
G^{-1}_{n n^{\prime }} (z ) \equiv 
(\omega_{n} -z )\delta_{n n^{\prime }} 
+ \lambda^2 s_{n n^{\prime }} (z ) ,
\label{eqn:4.20}
\end{equation}
which is the $(n, n^{\prime })$-th component of the $N \times N$ matrix 
$G^{-1} (z )$, and $s_{n n^{\prime}} (z)$ is defined by
\begin{equation}
s_{n n^{\prime}} (z) \equiv 
\int_{0}^{\infty}  d \omega^{\prime } 
\frac{ v_n (\omega^{\prime } )  v_{n^{\prime}}^* (\omega^{\prime} ) }
{z - \omega^{\prime }} , 
\label{eqn:3.60}
\end{equation}
for all $z=re^{ i  \varphi }$  
($r>0$, $0< \varphi < 2\pi$). 
Under the large-energy condition of Eq. (\ref{eqn:6.40}), 
$s_{n n^{\prime }} (z )$ is guaranteed to be 
analytic in the whole complex plane except the cut $[0, \infty )$. 
For the later convenience, 
$G^{-1} (z)$ is defined as an inverse of $G(z)$, 
where $G(z)$ is assumed to be regular. 
Note that $G(z)$ is nothing more than the reduced (or partial) resolvent 
$G_{n n^{\prime }}(z) = \langle n |(H -z)^{-1} | n^{\prime } \rangle$. 
One can confirm this fact by following 
the discussion in section 3.2 of Ref. \cite{Exner(1985)}. 
Since the behavior of $A(t)$ at long times is characterized 
by that of $F_n^{(\pm)} (\omega)$ in Eq. (\ref{eqn:220}) at small energies, 
we need to estimate the small-energy behavior of $G(z)$. 
Note that under the condition (\ref{eqn:6.40}) we have 
\begin{eqnarray}
G^{-1}_{n n^{\prime }} (\omega \pm i0) 
&=&
(\omega_{n} -\omega )\delta_{n n^{\prime }} 
\nonumber \\
&&
+ \lambda^2 
\bigl[ 
I_{n n^{\prime }} (\omega ) \mp i\pi v_n (\omega ) v_{n^{\prime }}^* (\omega ) 
\bigr]
\nonumber \\
&=&
\omega_n \delta_{n n^{\prime }} + \lambda^2 I_{n n^{\prime }} (0) +o(1) ,
\label{eqn:110}
\end{eqnarray}
as $\omega \rightarrow +0$, where 
$s_{n n^{\prime }} (\omega \pm i0) =
I_{n n^{\prime }} (\omega ) 
\mp i\pi v_n (\omega ) v_{n^{\prime }}^* (\omega )$ 
and 
$
I_{n n^{\prime}} (\omega )
\equiv 
P \int_0^{\infty} d\omega^{\prime} 
\frac{\varphi_n (\omega^{\prime} ) 
\varphi_{n^{\prime }}^* (\omega^{\prime} )}
{\omega -\omega^{\prime} } 
$, 
where $P$ denotes the principle value of the integral. 
The existence of $I_{n n^{\prime}} (0)$ 
may be just guaranteed by the small-energy condition of Eq. (\ref{eqn:6.40}). 
Supposing that $G_{n n^{\prime }} $ is of the form
\begin{equation}
G_{n n^{\prime }} (\omega \pm i0) =g_{n n^{\prime }} 
+o(1), 
\label{eqn:120}
\end{equation}
as $\omega \rightarrow +0$, 
one obtains that
\begin{eqnarray}
\delta_{n n^{\prime }} 
&=& 
\sum_{m=1}^{N}
G_{n m}  G^{-1}_{m n^{\prime }}  
\nonumber \\
&=&
\sum_{m=1}^{N}
g_{n m}
\bigl[
\omega_{m} \delta_{m n^{\prime }} + \lambda^2 I_{m n^{\prime }} (0)
\bigr] 
+o(1), 
\label{eqn:130}
\end{eqnarray}
which leads to
\begin{equation}
g_{n n^{\prime }} 
=\frac{1}{\omega_{n^{\prime }} }
\left[
\delta_{n n^{\prime }} - 
\lambda^2 \sum_{m=1}^{N} g_{n m} I_{m n^{\prime }} (0)
\right] .
\label{eqn:140}
\end{equation}
We solve this equation by assuming that 
$g_{n n^{\prime }} $ can be expanded for small $\lambda$ as
\begin{equation}
g_{n n^{\prime }} 
=
\sum_{j=0}^{\infty }
g_{n n^{\prime }}^{(j)} \lambda^{2j} .
\label{eqn:150}
\end{equation}
By substituting Eq. (\ref{eqn:150}) into (\ref{eqn:140}), it follows that 
\begin{equation}
g_{n n^{\prime }}^{(0)} 
=
\delta_{n n^{\prime }} /\omega_{n^{\prime }} 
,~~~
g_{n n^{\prime }}^{(1)} 
=
-I_{n n^{\prime }} (0)/\omega_n \omega_{n^{\prime }}  ,
\label{eqn:160}
\end{equation}
and for $j \ge 1$
\begin{equation}
g_{n n^{\prime }}^{(j)} 
=
-\frac{1}{\omega_{n^{\prime }}}
\sum_{m=1}^{N} g_{n m}^{(j-1)} I_{m n^{\prime }} (0) ,
\label{eqn:170}
\end{equation}
where we have assumed that all $\omega_n$ does not vanish. 
Note that 
$g_{n n^{\prime }}^{(0)} $ and $g_{n n^{\prime }}^{(1)} $ 
derived here accord with at least 
those for solvable cases, 
where $G(z)$ is explicitly obtained \cite{Antoniou(2004),Davies(1974)}. 
We can then obtain 
\begin{equation}
F_n^{(\pm)} (\omega )
=
-\lambda f_n \omega^p 
+o(\omega^p ) ,
\label{eqn:180}
\end{equation}
with
\begin{equation}
f_n \equiv
\frac{\tilde{q}_n }{\omega_n} 
-\lambda^2 
\sum_{n^{\prime} =1}^{N} 
\frac{I_{n n^{\prime }} (0) \tilde{q}_{n^{\prime }} }
{\omega_n \omega_{n^{\prime }} } +O(\lambda^4) ,
\label{eqn:240}
\end{equation}
where
\begin{equation}
\tilde{q}_n = 
\left\{ 
\begin{array}{ll}
q_n & ( p_n =p ) \\
0 & ( p_n \neq p )
\end{array} 
\right. ,
\label{eqn:230}
\end{equation}
where $p =\min \{ p_n \}$. 
With use of the $\tilde{q}_n$ instead of $q_n$, 
we extracted 
only the dominant part of $F_n^{(\pm)} (\omega )$ at small $\omega$.

The long time behavior of $A (t)$ can be simply obtained 
by applying to Eq. (\ref{eqn:6.20}) 
the asymptotic method for the Fourier integral \cite{Copson}. 
As mentioned before, 
the long time behavior is determined by the small-energy behavior of 
its integrand. 
By inserting Eq. (\ref{eqn:180}) 
into (\ref{eqn:220}), 
the integrand of $A (t)$ turns out to behave asymptotically 
\begin{equation}
|\langle \psi_{\omega }^{(\pm)} | \psi \rangle |^2
=
\lambda^2  \left| \sum_{n=1}^{N} f_n^* c_n  \right|^2 
\omega^{2p} +o(\omega^{2p}) ,
\label{eqn:6.100}
\end{equation}
as $\omega \rightarrow +0$. 
Applying the asymptotic formula for Fourier integrals, 
we obtain from Eq. (\ref{eqn:6.100}) 
the asymptotic behavior of Eq. (\ref{eqn:6.20}) reading,
\begin{equation}
A (t) = 
\lambda^2 \frac{\Gamma (2p+1)}{(it)^{2p+1}}
\left| \sum_{n=1}^{N} f_n^* c_n \right|^2 
+o(t^{-2p-1}) ,
\label{eqn:6.110}
\end{equation}
as $t \rightarrow \infty$, 
where $ i ^{d+1-p } =e^{ i (d+1-p )\pi /2}$, 
and $\Gamma(z+1)=\int_{0}^{\infty}  d x x^{z} e^{-x}$. 
We can clearly perceive $A(t) \sim t^{-2p-1} $, the power decay law.

Using the above result, let us first consider the higher-level effects 
on the long-time behavior that starts from 
the localized initial state at the lowest level. 
This study is directed to an examination of the SLA. 
For such an initial state, i.e., $c_n = \delta_{n 1}$, 
Eq. (\ref{eqn:6.110}) becomes 
\begin{equation}
A (t) = 
\lambda^2 \frac{\Gamma (2p+1)}{(it)^{2p+1}} 
\frac{|q_1 |^2 }{\omega_1^2} [1+ O(\lambda^2 ) ]
+o(t^{-2p-1}) ,
\label{eqn:8.13}
\end{equation}
where we supposed that $\tilde{q}_1 \neq 0$. 
Since there are no factors related to the higher levels 
in Eq. (\ref{eqn:8.13}), it implies that 
the long-time asymptotic behavior of $A (t)$ 
could agree with that in the SLA 
for a sufficiently small $\lambda$.

On the other hand, we can find a special superposition of 
unstable states $| n \rangle$ that 
maximizes the asymptote of $A(t)$ at long times. 
It is worth noting that 
its dependence on the initial states only appears in 
Eq. (\ref{eqn:6.110}) through the factor $\sum_{n=1}^{N} f_n^* c_n$, 
which can be rewritten by an inner product as
\begin{equation}
\sum_{n=1}^{N} f_n^* c_n
=
\langle \chi | \psi \rangle ,
\label{eqn:7.60}
\end{equation}
where we have introduced an auxiliary vector defined by 
\begin{equation}
| \chi \rangle 
\equiv 
\sum_{n=1}^{N} f_n | n \rangle .
\label{eqn:7.20}
\end{equation}
Thus, resorting to the Schwarz inequality, we see 
that the maximum of the factor (\ref{eqn:7.60}) 
is just attained by 
if and only if 
$| \psi \rangle \propto | \chi \rangle $, i.e., 
\begin{equation}
c_n = c f_n 
/ \|\chi \| ,
\label{eqn:7.70}
\end{equation}
where $c$ is an arbitrary complex number with $|c|=1$. 
Therefore, preparing the initial state $|\psi \rangle$ 
according to the above weights (\ref{eqn:7.70}), 
we can maximize the asymptote of $A(t)$ at long times. 
%
%
%
%
%
%
%
%
%
%
%
Substituting Eq. (\ref{eqn:7.70}) into Eq. (\ref{eqn:6.110}), 
one obtains that 
\begin{eqnarray}
A (t) 
&=& 
\lambda^2 \frac{\Gamma (2p+1) 
}{(it)^{2p+1}}
\| \chi \|^2 
+o(t^{-2p-1}) 
\label{eqn:8.25} \\
&\simeq& 
\lambda^2 \frac{\Gamma (2p+1)  
}{(it)^{2p+1}}
\sum_{n=1}^{N} \left| \frac{\tilde{q}_n }{\omega_n } \right|^2  
.
\label{eqn:260b}
\end{eqnarray}
It should be remarked that 
the initial state extended over unstable states $| n \rangle$ 
has the possibility of increasing the intensity of $A(t)$ 
more than a localized one would. 
This possibility may be interpreted as follows. 
Let us consider 
the spontaneous emission process for an atom 
interacting with the electromagnetic field, where 
$|n\rangle$ is identified with the $(n+1)$-th excited state of the atom 
with the vacuum state of the field 
and $|\omega \rangle$ is the ground state 
with the one-photon state. 
In this process, 
an initially excited atom makes a transition to the ground state 
with emitting a photon, 
while 
the atom that fell into the ground state 
can be reexcited by absorbing a photon. 
In the latter process, 
there are various candidates for the excited state. 
Repopulation of each excited level can make 
the intensity of $A(t)$ grow, 
providing that the initial state possesses those excited levels. 
However, 
if the initial state only consists of a specific excited state, 
the other excited states composing the state at a later time $t$ 
are discarded without any contribution to $A(t)$ \cite{comment}. 
This is the reason why the decay of the $A(t)$ 
for extended states can be relaxed more than that for localized states. 
%

Note that the above argument also suggests the possibility of finding 
another kinds of initial states that 
are coherently superposed to eliminate the factor (\ref{eqn:7.60}). 
This is indeed achieved by the initial states 
that are orthogonal to $| \chi \rangle $, 
\begin{equation}
\langle \chi | \psi \rangle 
=0 . 
\label{eqn:7.80}
\end{equation}
In this case, the first term in 
the rhs of Eq. (\ref{eqn:6.110}) becomes zero. 
This fact means that $A(t)$ for such an orthogonal state 
asymptotically decays faster than $t^{-2p-1}$.

The maximizing initial state 
seems to be desirable for an experimental verification 
of the power-decay law. 
Let us now discuss 
the value of $|A(t)|^2$ for such an initial state at long times. 
In particular, we shall evaluate this value at the time $t_{ep}$ 
of the transition from the exponential to the power decay law. 
We have to know the exact values of both $p_n$ and $q_n$ 
for many $n$'s; however, this requirement is satisfied 
by considering the spontaneous emission process for the hydrogen atom 
interacting with the electromagnetic field 
(see also Refs. \cite{Facchi(1998),Antoniou(2001)}). 
This time, $|n\rangle$ is interpreted as 
the $(n+1)p$-state of the atom 
with the vacuum state of the field, 
and $|\omega \rangle$ as the $1s$-state 
with the one-photon state. 
It then follows that $p_n =1/2$ for all $n$ \cite{Seke(1994)}, and 
$|q_n|$ is determined through the relation 
$\gamma_n = 2\pi \lambda^2 |v_n (\omega_n )|^2 +O(\lambda^2) 
\simeq 2\pi \lambda^2 |q_n|^2 |\omega_n|$, 
where the last estimation is confirmed in the dipole approximation. 
$\gamma_n$ is the decay rate of the $(n+1)p$-state, 
which is estimated as 
$\gamma_n 
\simeq 
8.0 \times 10^{9} \times 2^8 (n+1)n^{2n} / 9(n+2)^{2n+4} 
$ $\mathrm{s}^{-1}$ \cite{Bethe}. 
From these facts it follows that 
\begin{equation}
\lambda^2 \biggl| \frac{q_n}{\omega_n} \biggr|^2
\simeq 
\frac{8.0 \times 10^{9} \times 6 (n+1)^7 n^{2n} }
{\pi \Omega^3 (n+2)^{2n+4} [(n+1)^2 -1]^3 }
~\mathrm{s}^2 ,
\label{eqn:270}
\end{equation}
where $\omega_{ n} =\frac{4}{3}\Omega [1-(n+1)^{-2}]$ with 
$\Omega = 1.55 \times 10^{16}$ $\mathrm{s}^{-1}$, 
and we also choose $\lambda=6.43\times 10^{-9}$. 
Thus, we see that $|q_n|^2 / |\omega_n|^2 \sim n^{-3}$ for a large $n$. 
\begin{table}
\caption{
\label{tab:levels}
The level-number dependence of 
$\sum_{n=1}^{N} |q_n /\omega_n |^{2}$, 
the decay time $t_N$ of the $(N+1)p$-state, 
and the transition time $t_{ep}$ from the exponential to the power decay law. 
}
\begin{ruledtabular}
\begin{tabular}{cccc}
Number of levels $N$ &1&10&50 \\
\hline
$\bigl| \frac{\omega_1}{q_1} \bigr|^2 
\sum_{n=1}^{N} \bigl| \frac{q_n}{\omega_n} \bigr|^2 $ 
&1.00&1.28&1.29\\
%
%
$t_N ~(\mathrm{s})$&$1.60\times 10^{-9}$&$3.18\times 10^{-7}$
&$3.18\times 10^{-5}$ \\
$t_{ep}~(\mathrm{s})$&$2.00\times 10^{-7}$&$4.23\times 10^{-5}$
&$4.59\times 10^{-3}$ \\
%
%
%
\end{tabular}
\end{ruledtabular}
\end{table}
In Table \ref{tab:levels}, 
the numerical values of 
$
\sum_{n=1}^{N} |q_n /\omega_n |^{2}$ ($\simeq \| \chi \|^2 $), 
the decay time $t_N$ ($=1/\gamma_N$) of the $(N+1)p$-state, 
and 
the time $t_{ep}$ are listed 
for the three cases of the level numbers $N=1$, $10$, and $50$. 
%
%
Here, we define $t_{ep}$ as the maximum time which equates 
the square modulus of the asymptote (\ref{eqn:260b}) 
to that of the following $A(t)$ at intermediate times 
\cite{Antoniou(2003),Davies(1974)},
\begin{equation}
A(t) \simeq 
\sum_{n=1}^{N} |c_n|^2 e^{-it\omega_n -t\gamma_n /2} ,
\label{eqn:280}
\end{equation}
where $c_n$ is chosen as Eq. (\ref{eqn:7.70}). 
It is worth noting that when $t \gg t_N$, $|A(t)|^2$ can approximate 
$|c_N |^4 e^{-t\gamma_N}$ 
because the decay time $t_n$ lengthens with $n$ in this case. 
We see from Table \ref{tab:levels} 
that $t_{ep}$ is much longer than $t_N$, 
so that $t_{ep}$ is roughly estimated as the root of the equation, 
\begin{equation}
|c_N |^4 e^{-t\gamma_N} = 
\frac{\lambda^4 }{t^4}
\Biggl| \sum_{n=1}^{N} \biggl|\frac{q_n }{\omega_n} \biggr|^2 \Biggr|^2 . 
\label{eqn:290}
\end{equation}
On the other hand, the factor $\sum_{n=1}^{N} |q_n /\omega_n |^{2}$ 
is essentially unchanged with $N$, 
whereas $t_{ep}$ rapidly increases with $N$ (see Table \ref{tab:levels}). 
These facts and Eq. (\ref{eqn:290})  
imply that $A(t_{ep} )$ rather decreases as $N$ increases 
\cite{A(tep)}. 
Hence, we should unfortunately conclude that 
the maximizing initial state does not provide any help 
for an observation of the power decay law 
for the spontaneous emission from a hydrogen atom.

In summary, 
we have considered the long-time behavior 
of the unstable multilevel systems 
and estimated the asymptotic behavior of 
the survival amplitude $A(t)$ 
for an arbitrary initial state 
in the long-time region where $A(t)$ obeys a power decay law. 
We have then found two special initial states. 
One of them asymptotically maximizes $A(t)$ at long times, 
and the other eliminates the first term of the asymptotic expansion of $A(t)$. 
The latter fact may imply that 
the exponent of the power decay of $A(t)$ is determined by 
not only the small-energy behavior of the form factors 
but also the initial population in unstable states. 
%
%
Such relations between the initial states and the power decay law 
were studied with respect to the long-time behavior 
of wave packets, both for the free-particle system \cite{Miyamoto(free)} 
and for finite-range potential systems \cite{Miyamoto(potential)}. 
In the case of the experimental verification of the power decay laws, 
the existence of the maximizing initial states seems preferable. 
This expectation is probably misplaced 
for the spontaneous emission process of a hydrogen atom, 
however, a possibility still could remain for the systems 
allowing the photodetachment or the photoionization process. 
We then should require of them the property that 
both $\gamma_n$ and $|q_n /\omega_n |$ do not decrease as $n$ increases. 
More important states for this aim are those states 
which maximize $A(t)$ at the transition time 
from the exponential to the power decay law. 
%
%
The relation between such a maximizing state and the discussed one 
is still unclear. 
It will be addressed in a future issue.



The author would like to thank Professor I.\ Ohba 
and Professor H.\ Nakazato for useful comments, 
and Dr. J.\ Harada for helpful discussions. 
He would also like to thank the Yukawa Institute for Theoretical Physics 
at Kyoto University, where this work was initiated during 
the YITP-03-16, 
Quantum Mechanics and Chaos: From Fundamental Problems through Nanosciences. 
This work is partly supported by a Grant for The 21st Century COE Program 
at Waseda University from the Ministry of Education, Culture, 
Sports, Science and Technology, Japan. 




\begin{thebibliography}{99}



\bibitem{Nakazato(1996)}
For a review, see, for example, 
H.~Nakazato,\ M.~Namiki,\ and S.~Pascazio,  
\newblock  Int.\ J.\ Mod.\ Phys.\ B\ {\bf 10},\ 247\ (1996).



\bibitem{Khalfin(1957)}
L.~A.~Khalfin, \newblock Zh.\ Eksp.\ Theor.\ Fiz.\ 
{\bf 33},\ 1371 (1957)\ [Sov.\ Phys.\ JETP\ {\bf 6},\ 1053 (1958)]. 







\bibitem{Greenland(1988)}
P.~T.~Greenland, 
\newblock  Nature\ (London)\ {\bf 335},\ 298\ (1988).



\bibitem{Wilkinson(1997)}
S.~R.~Wilkinson,\ {\it et al}, 
\newblock  Nature\ (London)\ {\bf 387},\ 575\ (1997).




\bibitem{Friedrichs(1948)}
K.~O.~Friedrichs, 
\newblock Commun.\ Pure\ Appl.\ Math.\ {\bf 1},\ 361\ (1948). 




\bibitem{Exner(1985)}
P.~Exner, \newblock {\it Open Quantum Systems and Feynman Integrals}\ 
(Reidel,\ Doredrecht,\ 1985). 





\bibitem{Facchi(1998)}
P.~Facchi\ and S.~Pascazio,  
\newblock Phys.\ Lett.\ A\ {\bf 241},\ 139\ (1998). 



\bibitem{Antoniou(2001)}
I.~Antoniou,\ E.~Karpov,\ G.~Pronko,\ and E.~Yarevsky,  
\newblock Phys.\ Rev.\ A\ {\bf 63},\ 062110\ (2001). 



\bibitem{Rzazewski(1982)}
K.~Rz{\c a}{\. z}ewski,\ M.~Lewenstein,\ and J.~H.~Eberly,\ \newblock 
J.\ Phys.\ B\ {\bf 15},\ L661\ (1982). 




\bibitem{Haan(1984)}
S.~L.~Haan\ and\ J.~Cooper, \newblock J.\ Phys.\ B\ {\bf 17},\ 3481\ (1984).



\bibitem{Nakazato(2003)}
H. Nakazato, in ``Fundamental Aspects of Quantum Physics'', 
edited by L. Accardi and S. Tasaki (World Scientific, New Jersey, 2003).
















\bibitem{Frishman(2001)}
E.~Frishman\ and M.~Shapiro, 
\newblock Phys.\ Rev.\ Lett.\ {\bf 87},\ 253001\ (2001); 
\newblock Phys.\ Rev.\ A\ {\bf 68},\ 032717\ (2003). 







\bibitem{Antoniou(2003)}
I.~Antoniou,\ E.~Karpov,\ G.~Pronko,\ and E.~Yarevsky,  
\newblock Int.\ J.\ Theor.\ Phys.\ {\bf 42},\ 2403\ (2003). 





\bibitem{Antoniou(2004)}
I.~Antoniou,\ E.~Karpov,\ G.~Pronko,\ and E.~Yarevsky,  
\newblock  quant-ph/0402210\ (2004). 






\bibitem{Miyamoto(2003)}
M.~Miyamoto,\ 
in Proceedings of the Workshop on 
{\it Quantum Mechanics and Chaos: 
From Fundamental Problems through Nanosciences} (2003),\ 
to be appeared in Bussei\ Kenkyu\ (Kyoto) (in Japanese). 



\bibitem{Davies(1974)}
E.~B.~Davies, 
\newblock J.\ Math.\ Phys.\ {\bf 15},\ 2036\ (1974). 










\bibitem{Wigner(1948)} 
E.~P.~Wigner, \newblock Phys.\ Rev.\ {\bf 73},\ 1002\ (1948).




\bibitem{bound-states} 
This is the case for the strong-coupling interactions 
(see, e.g., \cite{Nakazato(2003)}), 
and for the degenerated multilevel systems \cite{Antoniou(2004)}. 





\bibitem{Copson}
E.~T.~Copson, \newblock 
{\em Asymptotic Expansions} 
\newblock (Cambridge Univ. Press,\ Cambridge,\ 1965),\ 
Chap.\ 3. 









\bibitem{comment}
Note that this interpretation is not necessarily 
applied to all of the time region. 
The reason is that such repopulation processes are naively expected to occur 
under the dominance of the terms of 
$O(\lambda^2)$ for our interaction Hamiltonian $\lambda V$. 
This does not contradict the behavior of $A(t)$ at long times 
\cite{Antoniou(2003)}. 




\bibitem{Seke(1994)}
J.~Seke,  
\newblock Physica\ A\ {\bf 203},\ 269\ (1994).



\bibitem{Bethe}
H.~A.~Bethe\ and E.~E.~Salpeter, 
\newblock {\em Quantum Mechanics of One- and Two-Electron Atoms} 
\newblock (Springer-Verlag,\ Berlin,\ 1957),\ 
Sect.\ 63. 





\bibitem{A(tep)}
For instance, we obtain, $|A(t_{ep})|^2 \simeq 4.41 \times 10^{-55}$ 
for $N=1$, an extremely small value. 






\bibitem{Miyamoto(free)} 
K.~Unnikrishnan, 
\newblock Am.\ J.\ Phys.\ {\bf 65},\ 526\ (1997);\ {\bf 66},\ 632 (1998); 
F.~Lillo\ and R.~N.~Mantegna, 
\newblock Phys.\ Rev.\ Lett.\  {\bf 84},\ 1061\ (2000);\ 
{\bf 84},\ 4516\ (2000);  
J.~A.~Damborenea,\ I.~L.~Egusquiza,\ and J.~G.~Muga, 
\newblock Am.\ J.\ Phys.\ {\bf 70},\ 738\ (2002); 
M.~Miyamoto, 
\newblock J.\ Phys.\ A\ {\bf 35},\ 7159\ (2002); 
\newblock Phys.\ Rev.\ A\ {\bf 68},\ 022702\ (2003). 




\bibitem{Miyamoto(potential)} 
M.~Miyamoto, \newblock Phys.\ Rev.\ A\ {\bf 69},\ 042704\ (2004).












\end{thebibliography}
\end{document}